\documentclass[11pt,a4paper]{article} 
\usepackage{jheppub}
\usepackage{mciteplus}

\def\gs{\mathrel{
   \rlap{\raise 0.511ex \hbox{$>$}}{\lower 0.511ex \hbox{$\sim$}}}}
\def\ls{\mathrel{
   \rlap{\raise 0.511ex \hbox{$<$}}{\lower 0.511ex \hbox{$\sim$}}}}

\newcommand{\be}{\begin{eqnarray}}
\newcommand{\ee}{\end{eqnarray}}

\def\be{\begin{equation}}
\def\ee{\end{equation}}
\newcommand{\ba}{\begin{array}{c}}
\newcommand{\baz}{\begin{array}{cc}}
\newcommand{\bad}{\begin{array}{ccc}}
\newcommand{\bav}{\begin{array}{cccc}}
\newcommand{\baf}{\begin{array}{ccccc}}
\newcommand{\bena}{\begin{eqnarray}}
\newcommand{\eena}{\end{eqnarray}}
\newcommand{\bea}{\begin{equation} \begin{array}{c}}
\newcommand{\eea}{ \end{array} \end{equation}}
\newcommand{\ea}{\end{array}}

\begin{document}

\subheader{\footnotesize\sc Preprint numbers: {ICCUB-12-477, YITP-SB-12-43}}
\title{Dark Radiation and Decaying Matter}

\author[a b]{M.~C.~Gonzalez-Garcia}
\author[c]{V.~Niro}
\author[d]{Jordi Salvado}

\affiliation[a]{Instituci\'o Catalana de Recerca i Estudis
  Avan\c{c}ats (ICREA),\\ 
  Departament d'Estructura i Constituents de la
  Mat\`eria and Institut de Ciencies del Cosmos,\\ Universitat de
  Barcelona, Diagonal 647, E-08028 Barcelona,
  Spain}

\affiliation[b]{C.N.~Yang Institute for Theoretical Physics, State University of New
  York at Stony Brook,\\ Stony Brook, NY 11794-3840, USA}

\affiliation[c]{Departament d'Estructura i Constituents de la Mat\`eria and Institut
  de Ciencies del Cosmos,\\ Universitat de Barcelona, Diagonal 647,
  E-08028 Barcelona, Spain}
\affiliation[d]{Wisconsin IceCube Particle Astrophysics Center (WIPAC) and Department of Physics,\\ 
University of Wisconsin, Madison, WI 53706, USA\\}

\emailAdd{concha@pheno0.physics.sunysb.edu}
\emailAdd{niro@ecm.ub.es}
\emailAdd{jordi.salvado@icecube.wisc.edu}

\abstract{Recent cosmological measurements 
favour additional relativistic energy density beyond the one provided by the
three active neutrinos and photons of the Standard Model (SM). This is often 
referred to as ``dark radiation'', suggesting the need of new light
states in the theory beyond those of the SM.  In this paper, we study
and numerically explore the alternative possibility that this increase
comes from the decay of some new form of heavy matter into the SM
neutrinos.  We study the constraints on the decaying matter density
and its lifetime, using data from the Wilkinson Microwave Anisotropy
Probe, the South Pole Telescope, measurements of the Hubble constant
at present time, the results from high-redshift Type-I supernovae and
the information on the Baryon Acoustic Oscillation scale.  We,
moreover, include in our analysis the information on the 
presence of additional contributions to the expansion rate of the Universe  
at the time of Big Bang Nucleosynthesis. 
We compare the results obtained in this decaying matter scenario 
with those obtained with the standard analysis in terms of a constant $N_{\rm
eff}$. }

\keywords{Cosmology of Theories beyond the SM, Neutrino Physics}

\maketitle

\section{\label{sec:intro} Introduction}
 
In recent years an enormous progress has been made in the field of
cosmology. Data on the Cosmic Microwave Background (CMB) has brought
cosmology into a precision science and have revealed a Universe made
by roughly 23\% of Dark Matter (DM) and 72\% of Dark Energy (DE).  For
a review on the physics of the CMB we refer to~\cite{Hu:2001bc}.
Furthermore, at present, recent analysis of cosmological data suggest a
trend towards the existence of ``dark radiation'' (see for example
\cite{Komatsu:2010fb,GonzalezGarcia:2010un,Hamann:2010bk,Nollett:2011aa,GonzalezMorales:2011ty,Joudaki:2012fx,Archidiacono:2012gv}).
Dark radiation alters the time of matter-radiation equality with a 
corresponding impact on the observed CMB 
anisotropies as well as an affect on the Large Scale Structure (LSS) 
distributions. 
The amount of dark radiation is usually parametrized using the
parameter $N_{\rm eff}$, that indicates the ``effective number'' of
neutrino-like relativistic degrees of freedom.  The value associated
with the standard case of three active neutrino flavours was
calculated in detail in Ref.~\cite{Mangano:2005cc} and was found to be
$N^{\rm SM}_{\rm eff}=3.046$.

The Wilkinson Microwave Anisotropy Probe (WMAP) collaboration found
$N_{\rm eff} = 4.34^{+0.86}_{-0.88}$ based on their 7-year data
release and additional LSS 
data~\cite{Komatsu:2010fb} at $1 \sigma$ in a $\Lambda$CDM cosmology.
More recent measurements of the CMB anisotropy on smaller
scales by the Atacama Cosmology Telescope (ACT)~\cite{Das:2010ga} and
South Pole Telescope (SPT)~\cite{Keisler:2011aw} experiments seem to also
favour a value of $N_{\rm eff}$ higher than predicted in SM. 
In a $\Lambda$CDM cosmology  the current constraints on $N_{\rm eff}$ 
at 68\%~C.L at CMB time read~\cite{Abazajian:2012ys}:  
\bena 
N^{\rm CMB}_{\rm eff}=4.34^{+0.86}_{-0.88}\quad&\text{ WMAP7+BAO+}H_0\,,\\
N^{\rm CMB}_{\rm eff}=3.86 \pm 0.42 \quad&\text{ WMAP7+SPT+BAO+}H_0\,,
\label{eq:SPT}\\
N^{\rm CMB}_{\rm eff}=3.89 \pm 0.41 \quad&\text{
WMAP7+ACT+SPT+BAO+}H_0\,,
\label{eq:ACTSPT}
\eena 
where $H_0$ refers to the constraint $H_0= 74.2 \pm
3.6$~km~s$^{-1}$ found by the Hubble Space
Telescope~\cite{Riess:2009pu}.  This evidence of dark radiation is robust
under consideration of more generalized  cosmologies. For example 
in Ref.~\cite{GonzalezGarcia:2010un}, $N_{\rm eff} =
4.35^{+1.4}_{-0.54}$ was found in a global analysis including the data
from cosmic microwave background experiments (in particular from WMAP-7), the Hubble constant $H_0$ measurement \cite{Riess:2009pu},
the high-redshift Type-I supernovae\cite{Hicken:2009df} and the LSS
results from the Sloan Digital Sky Survey (SDSS) data release 7 (DR7)
halo power spectrum~\cite{Reid:2009xm}. The value refers to generalized cosmologies
which depart from $\Lambda$CDM models by allowing not only the
presence of dark radiation but also dark energy with equation of state
with $\omega\neq -1$, neutrino masses, and non-vanishing
curvature. 

Independent information on the amount of radiation at earlier times
is provided by its effect on the  expansion rate of the Universe 
at the time of Nucleosynthesis. Faster expansion would lead to an earlier 
freeze-out of the neutron to proton ratio and would lead to a higher $^4$He
abundance generated during Big Bang Nucleosynthesis (BBN)
\cite{Yang:1978ge}.  Consequently the value of $N_{\rm eff}$ at the time 
of BBN can be constrained using primordial Nucleosynthesis yields of
deuterium and helium.  Old data and analysis reported a value of
$N_{\rm eff}$ at BBN consistent with the standard model prediction:
$N^{\rm BBN}_{\rm eff}=2.4 \pm 0.4$ at 68\%~C.L.~\cite{Simha:2008zj}.
However recent results indicate, as well, that the
relatively high $^4$He abundance can be interpreted in terms of
additional radiation during the BBN
epoch~\cite{Izotov:2010ca,Aver:2010wq,Mangano:2011ar} (see
\cite{Steigman} for a recent review).  
In particular in Ref.~\cite{Mangano:2011ar} it was found
\begin{equation}
N^{\rm BBN}_{\rm eff}
=3.83^{+0.22}_{-0.61}
\label{eq:bbnneff}
\end{equation}
at 68\%~C.L. which  suggests that $\Delta
N^{\rm BBN}_{\rm eff} < 1$ at 95\%~C.L. at BBN time. 
This bound is reasonably independent of measurements on the baryon density 
from CMB anisotropy data and of the neutron lifetime input. 

The number of active neutrinos is constrained by measurements of the
decay width of the Z~boson~\cite{Nakamura:2010zzi} to be~$2.984 \pm
0.008$.  For this reason, several authors invoked the presence of a
sterile neutrino to explain the above
data~\cite{Acero:2008rh,Hamann:2010bk,Giusarma:2011zq,Archidiacono:2011gq,Archidiacono:2012ri}.
Furthermore, the possibility of one or two additional eV scale mass
sterile neutrinos, brought to equilibrium with the SM ones via their
mixing, could also account for some of the anomalies observed in
short-baseline (SBL) neutrino experiments, that favour one or two
sterile
neutrinos~\cite{Kopp:2011qd,Giunti:2011gz,Giunti:2011hn,Giunti:2011cp,Karagiorgi:2012kw,Donini:2012tt}.

Light sterile neutrinos account for dark radiation in the same amount
at BBN and CMB times, {\it{i.e.}} in these scenarios $N^{\rm BBN}_{\rm
eff}=N^{\rm CMB}_{\rm eff}$. Alternatively and in the light of the old
BBN value for $N_{\rm eff}$~\cite{Simha:2008zj}, other models were
proposed to explain an increasing on $N_{\rm eff}$ at CMB time, while
having $N_{\rm eff}$ compatible with three at BBN time. In
Ref.~\cite{Fischler:2010xz}, for example, the authors suggested that
the increase in radiation could be explained by the 
decay of non-relativistic matter into relativistic states -- see
also~\cite{Menestrina:2011mz,Ichikawa:2007jv, Choi:2012zn} for a
general study on dark radiation production from particle decays --
while, 
in Ref.~\cite{Hooper:2011aj}, it was proposed that non-thermal
DM production, such as late-time decays of a long-lived state into DM and
neutrinos and photons, could mimic an additional neutrino species. 
Recently, in Ref.~\cite{Boehm:2012gr}, the authors discussed the possibility 
that a new particle of mass $\lesssim$~10~MeV, that remains in thermal
equilibrium with neutrinos until it becomes non-relativistic, leads to
a value of $N_{\rm eff}$ that is greater than three. 

In this paper, we explore the possibility of generating the dark
radiation without the need of additional light states besides the ones
of the SM. In particular, we consider a scenario in which dark
radiation consists of SM neutrinos which have been populated in an
extra amount by the decay of some heavy form of thermally produced
matter as described in Sec.~\ref{sec:frame}.  One important constraint
in our study is the requirement that the decay products, which
constitute the dark radiation, are SM neutrinos. This is so because
before neutrino decoupling any neutrinos produced by the decay would
be brought to equilibrium by SM interactions and, in this way, the
additional contribution from dark radiation to the expansion would be
erased.

We perform a global analysis of the relevant cosmological observables 
to determine the allowed range for the neutrino decaying matter density and 
its lifetime in this framework. Our results, presented in 
Sec.~\ref{sec:results}, show that this scenario can explain the tendency of 
the data to favour more radiation 
at the BBN and CMB times without the need of adding one or more sterile
neutrinos nor any other new relativistic states to the three active ones of 
the SM. In particular we find that a lifetime of order 
$\tau_{\rm dec}\sim 10^3$~s is favoured, implying a decay that happens 
during BBN time. We summarize our conclusions in Sec.~\ref{sec:conclu}.

\section{\label{sec:DM_dec} Decaying Matter}
\label{sec:frame}
Generically when discussing the possibility of decay matter as the
origin of dark radiation one can think of two possibilities:
in the first scenario the decaying matter is the main component of 
the DM itself, while in the second case the particle that decays is not 
the relic particle present nowadays in the Universe, but it 
is another particle species.

The possibility of decaying DM (DDM) has been exhaustively explored 
in the literature. CMB constraints on decaying warm DM have 
been derived in~\cite{Lattanzi:2007ux}, specifically in the context of 
a Majoron DM. Cosmological constraints on the DM decay rate into neutrinos have
been obtained in Ref.~\cite{Gong:2008gi}, using only Type Ia supernova
data (SNIa) and the first CMB peak, see
also~\cite{Ichikawa:2008pz,Ichikawa:2006vm,Aoyama:2011ba,Wang:2012ek}
for more references.  More recently, in
Ref.~\cite{DeLopeAmigo:2009dc}, the authors extended the analysis
using the full CMB anisotropy spectrum, large scale structure (LSS),
Lyman-$\alpha$ data and weak lensing observations: the authors found a
DM lifetime of the order $\Gamma^{-1}_{dec} \gtrsim
100$~Gyr. Cosmological data, thus, tightly constrain the DM decay rate
into neutrinos. This happens because DDM alters the time of
matter-radiation equality, and the early integrated Sachs-Wolfe
effect, with an increase in the first CMB peak.  DDM also changes the
late integrated Sachs-Wolfe effect, with a direct consequence on the
CMB anisotropy spectrum at small multipoles,
see~\cite{Lattanzi:2007ux} for a detailed explanation.  Bounds on the
DM lifetime can be obtained independently also from neutrino
telescopes data.  The Super-Kamiokande bounds on the DM flux from the
Galactic Center, from the Earth and the Sun are given
in~\cite{Desai:2004pq} and, using these data, the constrain on the DM
lifetime has been calculated in
Refs.~\cite{PalomaresRuiz:2007ry,Covi:2009xn}.

Consequently since the DM decay rate into neutrinos is constrained to be much
longer than the age of the Universe, the increase in the value of
$N_{\rm eff}$ that we expect in the DDM scenario is, in general, tiny
\footnote{Considering a DM lifetime that depends on time, as suggested
in Ref.~\cite{Bjaelde:2012wi}, it is possible to explain the excess in
the radiation within the DDM scenario.}.  

Alternatively one can consider scenarios -- denoted in the following
as $dec$M -- in which the decaying state is different that the one
dominantly present nowadays in the Universe as dark matter, such as
those in Ref.\cite{Ichikawa:2007jv,Fischler:2010xz,Menestrina:2011mz,Hooper:2011aj,Choi:2012zn}. In what
follows, we will consider that the $dec$M is thermally produced and
dominantly decays into neutrinos. 
We, moreover, restrict our study to the case of a non-relativisti $dec$M at BBN time: its mass is greater 
than roughly 10~MeV.

\subsection{\label{sec:BE} Boltzmann Equations for Decaying Matter
}

In the synchronous gauge, the line element is defined as: \be
ds^2=a^2(\tau)\{-d\tau^2+(\delta_{ij}+h_{ij})dx^i dx^j\}\,, \ee where
$\tau$ is the conformal time, while $t$ is the cosmological time
($dt=a(\tau)\,d\tau$), and $a=R/R_0$ is the scale factor. The metric
perturbation $h_{ij}$ can be expanded in Fourier space
as~\cite{Ma:1995ey} \be h_{ij}(\vec{x},\tau)=\int d^3k
e^{i\vec{k}\cdot\vec{x}} \{\hat{k}_i \hat{k}_j
h(\vec{k},\tau)+(\hat{k}_i \hat{k}_j - \frac{1}{3} \delta_{ij}) 6
\eta(\vec{k},\tau)\}\,, \ee with $\vec{k}=k \hat{k}$. We will use the
notation defined above when writing the equations for the
perturbations, i.e. the fields $h(\vec{k},\tau)$ and
$\eta(\vec{k},\tau)$.

In the $dec$M scenario, the set of Boltzman equations 
describing the background evolution of the different components of the 
energy density of the Universe is complemented with two 
equations describing  the evolution of the $dec$M and dark radiation 
density as~\cite{Scherrer:1984fd,Scherrer:1987rr}: 
\bena
\dot{\rho}_{dec} & = & - 3 a H \rho_{dec} - a
\Gamma_{dec}
\,\rho_{dec}\,,
\label{eq:rho_DM}\\ 
\dot{\rho}_{dr} & = &
-4 a H \rho_{dr} + a
\Gamma_{dec}\,\rho_{dec}\,,\label{eq:rho_DR}
\label{eq:rho_SR}
\eena where the over-dot denotes the derivative respect to conformal
time $\tau$.  The subscript ``$dr$'' indicates the dark radiation
coming from the $dec$M, while ``$dec$'' is the density of the $dec$M. 
$\Gamma_{dec}$ is the $dec$M decay width.

Correspondingly we derive the equations for the density 
fluctuations~\cite{Ma:1995ey}  generalized for the case of $dec$M 
into SM neutrinos by adapting the equations for of DDM 
in~\cite{Kaplinghat:1999xy,Ichiki:2004vi,Lattanzi:2008ds} to the
specifics of our case, in which the decaying products are SM neutrinos.
We find:
\bena 
\dot{\delta}_{\scriptscriptstyle DM} & = & -\frac{1}{2}
\dot{h}\,,\\ 
\dot{\delta}_{\scriptscriptstyle R} & = & -\frac{2}{3}
\dot{h} -\frac{4}{3} \theta_{\scriptscriptstyle R} + a \Gamma_{dec}
\frac{\rho_{dec}}{\rho_{\scriptscriptstyle
R}}\left(\delta_{\scriptscriptstyle DM}-\delta_{\scriptscriptstyle
R}\right)\,,\\ 
\dot{\theta}_{\scriptscriptstyle R} & = & k^2
\left(\frac{1}{4} \delta_{\scriptscriptstyle R} -
\sigma_{\scriptscriptstyle R}\right) - a \Gamma_{dec}
\frac{\rho_{dec}}{\rho_{\scriptscriptstyle R}}
\theta_{\scriptscriptstyle R}\,,\\ 
\dot{\sigma}_{\scriptscriptstyle R}
& = & \frac{1}{2}\left(\frac{8}{15} \theta_{\scriptscriptstyle R} -
\frac{3}{5} k F_{3} + \frac{4}{15} \dot{h} + \frac{8}{5} \dot{\eta}
\right) - a \Gamma_{dec} \frac{\rho_{dec}}{\rho_{\scriptscriptstyle
R}} \sigma_{\scriptscriptstyle R}\,,\\ 
\dot{F}_{l} & = &
\frac{k}{2l+1} \left[l F_{l-1} - \left(l+1\right) F_{l+1}\right] - a
\Gamma_{dec} \frac{\rho_{dec}}{\rho_{\scriptscriptstyle R}} F_{l}\,,
\label{eq:Fl}
\eena 
where we have
defined $\rho_{\scriptscriptstyle R}\equiv \rho_\nu+\rho_{dr}$,
$\rho_{\scriptscriptstyle DM} \equiv \rho_{c}+\rho_{dec}$, with
$\rho_\nu$ the standard neutrino contribution, i.e. non coming from
matter decay, and $\rho_{c}$ the density of standard cold dark matter.
In Eq.~(\ref{eq:Fl}) $l\geq3$ and $F_2=2 \sigma_{\scriptscriptstyle
R}$ and we have used the conventions in
Refs.~\cite{Ma:1995ey,DeLopeAmigo:2009dc}. Note that the
density perturbation equation for the DM does not depend on the value
of $\Gamma_{dec}$ since both the background DM density and
overdensity decay at the same rate~\cite{DeLopeAmigo:2009dc}. 

We have implemented the previous equations for the
background and the perturbations in the CLASS (Cosmic Linear
Anisotropy Solving System) code~\cite{Blas:2011rf}. We considered flat
cosmology and an equation of state for the cosmological constant
parameter given by $\omega=-1$.

\section{\label{sec:num} Numerical Analysis}

\subsection{\label{sec:num:inputs} Cosmological Inputs}

In our analysis we include the results from the 7-year data of
WMAP~\cite{Komatsu:2010fb} on the temperature and polarization
anisotropies, using the likelihood function as provided by the
collaboration.  The ACT~\cite{Dunkley:2010ge} and
SPT~\cite{Keisler:2011aw} experiments have probed higher multipole
moments than WMAP. We implement the SPT data only in our analysis,
since they give a bound in $N_{\rm eff}$ of the same order as the one
obtained considering both the ACT and the SPT data, see
Eqs.~\eqref{eq:SPT},~\eqref{eq:ACTSPT}. We build the corresponding
likelihood functions from the data, covariance matrix and window
functions, introducing other three parameters in the analysis: the
Sunyaev-Zel'dovich (SZ) amplitude, the amplitude of Poisson
distributed point sources and the amplitude of clustered point
sources. For these parameters, we used gaussian priors as given in
Refs.~\cite{Shirokoff:2010cs,Keisler:2011aw} and the templates of
Ref.~\cite{Keisler:2011aw}. We fixed the foreground terms to be
positive.  In the following, with the notation ``CMB'' we will always
refer to the combination of WMAP7 and SPT data. 

We introduce a Hubble parameter prior, based on the latest Hubble
Space Telescope value~\cite{Riess:2011yx}: $H_0=73.8 \pm
2.4$~km~s$^{-1}$~Mpc$^{-1}$.  This measurement of $H_0$ is obtained
from the magnitude-redshift relation of 240 low-z Type Ia supernovae
at $z < 0.1$.

We also include the luminosity measurements of high-z SNIa as given in
Ref.~\cite{Hicken:2009df}.  This compilation, the ``Constitution''
set, consists of 397 supernovae and it is an extension of the previous
sample, the ``Union'' set~\cite{Kowalski:2008ez}.

Finally we use the measurement of BAO scale obtained from the
Two-Degree Field Galaxy Redshift Survey (2dFGRS) and the Sloan Digital
Sky Survey Data Release 7 (SDSS DR7)~\cite{Percival:2009xn}.  We use
as input data the two distance ratios $d_z$ at $z = 0.2$ and $z =
0.35$, with $d_z \equiv r_s(z_d)/D_V(z)$, where $r_s(z_d)$ is the
comoving sound horizon at the baryon drag epoch and $D_V (z) = [(1 +
z)^2 D^2_A c z / H(z)]^{1/3}$, with $D_A$ the angular diameter
distance and $H(z)$ the Hubble parameter. We build the corresponding
likelihood function using the covariance matrix as given in
Ref.~\cite{Percival:2009xn}.  Since in the fitting procedure
of~\cite{Percival:2009xn}, the value of $d_z$ is obtained by first
assuming some fiducial cosmology ($h=0.72$, $\Omega_b h^2=0.0223$,
$\Omega_m=0.25$), we rescale the predictions for the BAO scale as
explained in Ref.~\cite{Hamann:2010pw}.  We do not introduce the
information on the full power-spectrum of the SDSS DR7
survey~\cite{Reid:2009xm}, but we consider only the BAO measurement,
since we will neglect the neutrino mass in our analysis.

The analysis method we adopt is based on a Markov Chain Monte Carlo
(MCMC) generator which employs the Metropolis-Hasting algorithm,
see~\cite{GonzalezGarcia:2010un,GonzalezGarcia:2009ya} for more
details.  For convenience for the case of $dec$M  in the  
MC generator we use the parameters in 
the  Table \ref{tab:definitions} for which we assume a
flat prior.  
The parameters $n_s, \tau, A_{\scriptscriptstyle S}$ and
$\Gamma_{dec}/\textrm{Gyr}^{-1}$ are respectively the scalar
spectral index, the optical depth at reionization, the amplitude of
scalar power spectrum at $k=0.05$~Mpc$^{-1}$ and the
decay rate into neutrinos.  

The first four parameters in Table~\ref{tab:definitions}  
are related to the initial
cosmological constant, baryon, DM and $dec$M density as:
\be
\rho^i_\Lambda = \tilde{\Omega}_\Lambda \cdot \tilde{H}^2_0
\ee
\be
\rho^i_b = \tilde{\Omega}_b \cdot \tilde{H}^2_0 \cdot a^{-3}
\ee
\be
\rho^i_c = \tilde{\Omega}_c \cdot \tilde{H}^2_0 \cdot a^{-3}
\ee
\be
\rho^i_{dec} = \tilde{\Omega}_{dec} \cdot \tilde{H}^2_0 \cdot a^{-3}
\ee
where $\tilde{H}^2_0$ has a fixed value. The densities are all in
units of $[3c^2/8\pi G]$, with $c$ the speed of light and $G$ the
Newton's constant.  Note that we have decided to use 
$\tilde{\Omega}_\Lambda$ as MC parameter rather than $\tilde{H}^2_0$
and that the parameter $\tilde{H}^2_0$ is not equivalent to the Hubble
parameter today.  The ``physical'' Hubble parameter is calculated at
each redshift using the Friedmann equation: 
\be H(z)= \sqrt{\sum_i
\rho_i (z)}\,, 
\ee 
with $i$ that runs over all the components of the Universe. 
After solving the Boltzmann equations, we obtain the present 
values ($z=0$) of the Hubble constant $H_0$, of the baryon density $\Omega_b$ 
and of the cold dark matter density $\Omega_c$. More specifically, 
the latter parameter is defined as $\Omega_c=(\rho_c(z=0)+\rho_{dec}(z=0))/\rho_{crit}$, 
but we will see that our results favour a negligible contribution of $\rho_{dec}$ 
to the dark matter density at present time. 
When presenting the results of our analysis, we will use 
these derived  parameters (which we refer to as ``analysis parameters'') 
to better compare with the  $\Lambda$CDM+$N_{\rm eff}$ case.  

At any time (or equivalently redshift $z$) the densities 
$\rho_{dec}(z)$ and $\rho_{dr}(z)$ can be translated in a time dependent effective number 
of neutrinos depending on how this one is related with the observations.
At BBN $N_{\rm eff}$ is determined by its contribution to the expansion
rate of the Universe which is what affects the primordial abundances. 
Any form of energy density, relativistic or non-relativistic enters 
in this observable. Consequently, at BBN  
\be
\Delta N_{\rm eff}^{\rm BBN}= 
\frac{\rho_{dec}(z_{\rm BBN})+\rho_{dr}(z_{\rm BBN})}
{\left( \frac{7}{8}\right) \left( \frac{4}{11}\right)^{4/3} 
\rho_\gamma(z_{\rm BBM})}\equiv
N_{\rm eff}^{\rm BBN}-N^{\rm SM}_{\rm eff}\,,
\ee
with $N^{\rm SM}_{\rm eff}=3.046$. 
To account for the constraints from BBN we impose
in the analysis a prior on the value of $N^{\rm BBN}_{\rm eff}$, 
Eq.~(\ref{eq:bbnneff}) defined for definiteness, at the time at which BBN 
ends, $z_{\rm BBN}=T_{\rm BBN}/T_\gamma^0-1$ with  $T_{\rm BBN}=0.01$~MeV
($T^0_\gamma=$2.726~K).

In CMB observables, the information on $N_{\rm eff}$ arises mostly from its 
contribution to the determination of the the matter-radiation equality epoch. 
Thus, when translating our results in terms of 
$N_{\rm eff}$ at CMB time, we will define
\be
\Delta N_{\rm eff}^{\rm CMB}= 
\frac{\rho_{dr}(z_{\rm CMB})}
{\left( \frac{7}{8}\right) \left( \frac{4}{11}\right)^{4/3} 
\rho_\gamma(z_{\rm CMB})}\,,
\ee
with $z_{\rm CMB}=1100$.

Furthermore, to implement thermal equilibrium condition on the 
dark radiation before neutrino decoupling, we impose
\begin{equation}
\frac{\rho_{dr}(z_{\rm \nu-decoup})+\rho_{\nu}(z_{\rm \nu-decoup})}
{\left( \frac{7}{8}\right) \left( \frac{4}{11}\right)^{4/3} 
\rho_\gamma(z_{\rm \nu-decoup})}=3.046\; ,
\label{eq:decneff}
\end{equation}
with $T_{\rm \nu-dec}\geq 1$~MeV .

Using the above data and their theoretical predictions -- obtained
from the modified Boltzmann equations presented in the previous section -- 
we construct the combined likelihood function and the corresponding    
probability distribution function from which we obtain the  
one-dimensional and two-dimensional probability distributions, as
described in Ref.~\cite{GonzalezGarcia:2010un}.

For the sake of comparison, we also perform  the
corresponding analysis in the framework of a $\Lambda$CDM model with a constant
$\Delta N_{\rm eff}$ (but without imposing the condition Eq.(\ref{eq:decneff})).
In this case, the parameters used in the MC, for which flat priors 
are used, are the present Hubble constant 
$H_0$, baryon density $\Omega_b$, cold dark matter density $\Omega_c$ 
and $\Delta N_{\rm eff}$, together with $n_s, \tau$, and  
$A_{\scriptscriptstyle S}$. 

\begin{table}[!t]
\centering
\begin{tabular}{l c}
\hline
MC Parameters & symbols \\ [0.5ex]
\hline \hline
Cosmological constant density & $\tilde{\Omega}_\Lambda$\\
Baryon density & $\tilde{\Omega}_b$\\
Dark Matter density & $\tilde{\Omega}_c$\\
Decaying Matter density & $\tilde{\Omega}_{dec}$)\\
Scalar spectral index & $n_s$\\
Optical depth at reionization & $\tau$\\
Amplitude of scalar power spectrum at $k=0.05$~Mpc$^{-1}$ 
& $A_{\scriptscriptstyle S}$ \\ 
Decay rate & $\Gamma_{dec}/\textrm{Gyr}^{-1}$\\ 
\hline
\end{tabular}
\caption{Parameters used in the MC generation in the $dec$M scenario.}
\label{tab:definitions}
\end{table}

\subsection{\label{sec:results} Results}

\begin{table}[!t]
\centering
\begin{tabular}{|l || c | c | c | c | c | c | c | c | c |}
\hline
& \multicolumn{3}{|c|}{$dec$M+BBN}& 
\multicolumn{3}{|c|}{$\Lambda$CDM+$N_{\rm eff}$+BBN}\\ \hline
Parameter & best & 1$\sigma$ & 95\% &
 best & 1$\sigma$ & 95\% \\ [1ex]
\hline
$H_0$ [km/s/Mpc] & 72.6 &  $^{+1.5}_{-1.4}$  & $^{+3.0}_{-2.8}$ 
& 73.0 &  $^{+1.5}_{-1.5}$  & $^{+2.8}_{-3.0}$\\[6pt]
$\Omega_b h^2 \times 100$ & 2.254&  $^{+0.034}_{-0.037}$  & $^{+0.069}_{-0.068}$ 
& 2.258&  $^{+0.032}_{-0.037}$  & $^{+0.065}_{-0.070}$ \\[6pt]
$\Omega_c h^2$  & 0.125&  $^{+0.005}_{-0.005}$  & $^{+0.012}_{-0.010}$ 
& 0.127 &  $^{+0.006}_{-0.006}$  & $^{+0.011}_{-0.012}$\\[6pt]
$\log(\rho_{dec}/\rho_\gamma)$ at $t=10^{-4}$~s & -4.61&  $^{+0.61}_{-0.73}$  & $^{+0.92}_{-1.7}$ 
& -- & -- & --\\[6pt]
$n_s$   & 0.973&  $^{+0.009}_{-0.009}$  & $^{+0.018}_{-0.018}$ 
& 0.975&  $^{+0.010}_{-0.010}$  & $^{+0.019}_{-0.019}$ \\[6pt]
$\tau$ & 0.084&  $^{+0.013}_{-0.015}$  & $^{+0.026}_{-0.026}$ 
& 0.083&  $^{+0.013}_{-0.013}$  & $^{+0.027}_{-0.024}$\\[6pt]
$A_s \times 10^9$  & 2.452&  $^{+0.082}_{-0.083}$  & $^{+0.164}_{-0.157}$ 
& 2.449&  $^{+0.075}_{-0.083}$  & $^{+0.155}_{-0.159}$\\ [6pt]
$\log(\tau_{dec}/\rm{s})$ & 2.9&  $^{+1.7}_{-1.0}$  & $^{+3.7}_{-1.5}$ 
& -- & -- & --\\[6pt]
$\Delta N_{\rm eff}^{\rm CMB}$ & 0.50&  $^{+0.30}_{-0.19}$  & $^{+0.58}_{-0.42}$ 
& 0.70&  $^{+0.25}_{-0.30}$  & $^{+0.44}_{-0.59}$\\[6pt]
$\Delta N_{\rm eff}^{\rm BBN}$ & -- & -- & $\leq 0.90$ 
& 0.70&  $^{+0.25}_{-0.30}$  & $^{+0.44}_{-0.59}$\\ [1ex]\hline
\end{tabular}
\caption{Best-fit values, 68\%~C.L. and 95\%~C.L. errors for the 
analysis parameters.}
\label{tab:results}
\end{table}

\begin{figure}[!t]
\centering
\includegraphics[width=14cm]{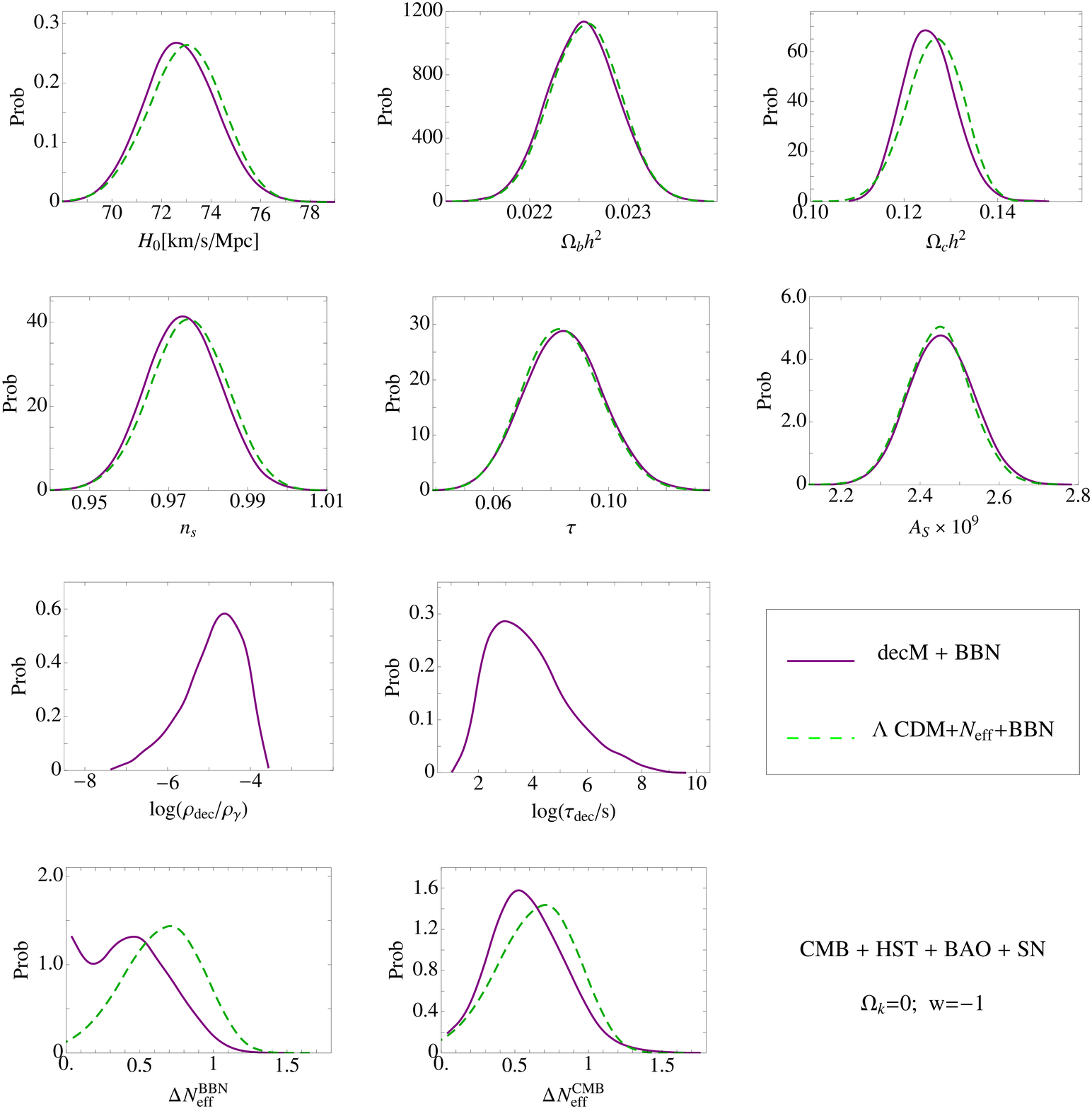}
\caption{Marginalized one-dimensional probability distributions for
the analysis parameters. We
present the results for the case of $dec$M+BBN with solid lines. 
The dashed line shows the results for $\Lambda$CDM+$N_{\rm eff}$+BBN. }
\label{fig:results1D}
\end{figure}

\begin{figure}[!t]
\includegraphics[width=14cm]{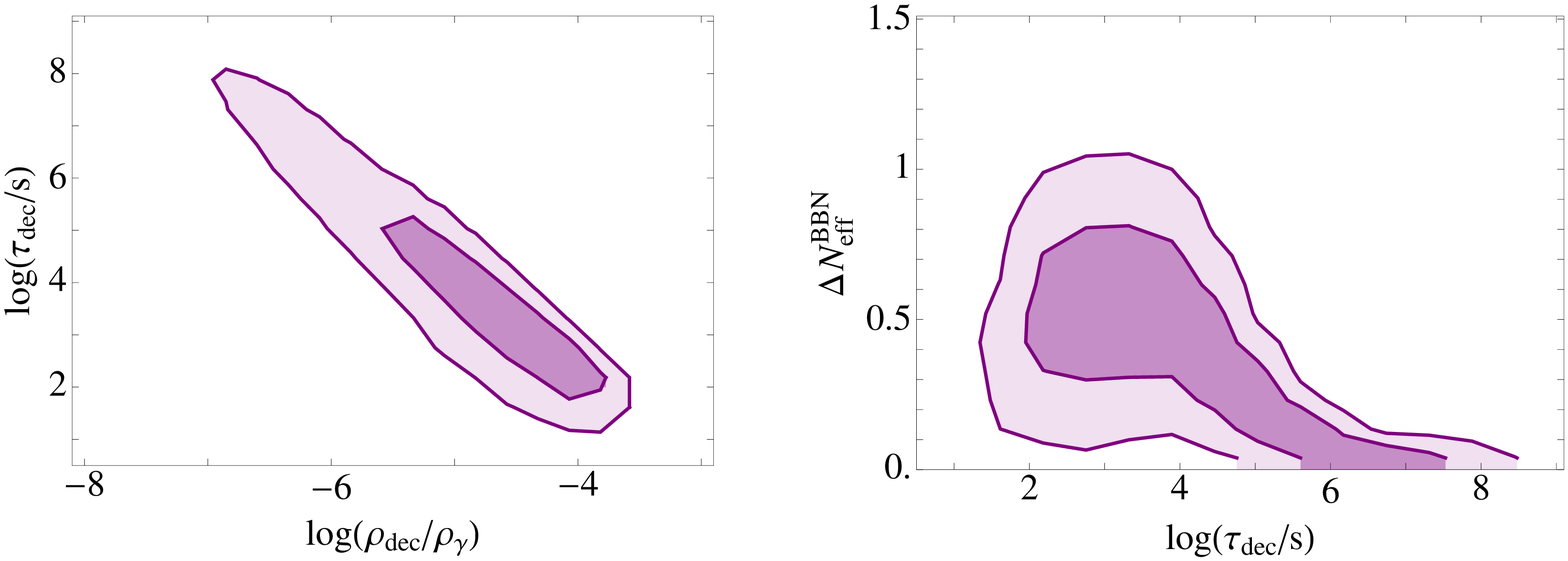}
\caption{Two-dimensional probability distribution at the 68\% and
95\%~C.L. for the $dec$M parameters $\{\log(\rho_{dec}/\rho_\gamma)$,
$\log(\tau_{dec}/\rm{s}) \}$, and $\{\log(\tau_{dec}/\rm{s})$, $\Delta
N_{\rm eff}^{\rm BBN}\}$.} 
\label{fig:results2D_dec}
\end{figure}

The results of our cosmological fits are summarized in
Tab.~\ref{tab:results} and in Figs.~\ref{fig:results1D}
and~\ref{fig:results2D_dec}.  In Tab.~\ref{tab:results}, we report the
best-fit values as well as the $1\sigma$ and 95\%~C.L. range for the
analysis parameters. We present the results for our model, the $dec$M
case, as well as for the standard $\Lambda$CDM model with the addition
of a fix $N_{\rm eff}$ and with the implementation of the constraint on
$N^{\rm BBN}_{\rm eff}$ in both cases.  We denote these two analysis
as $dec$M+BBN and $\Lambda$CDM+$N_{\rm eff}$+BBN. As seen from the
table, in the $dec$M+BBN the set of data considered is
better described with an amount of decaying matter
with lifetime short enough to allow for a good fraction of it to have
decayed into neutrinos at BBN time.

In Fig.~\ref{fig:results1D}, we present the one-dimensional probabilities
for the two analysis. For a better understanding, we report both the
value of $\Delta N_{\rm eff}$ at BBN and CMB time (which for
$\Lambda$CDM+$N_{\rm eff}$+BBN are identical by definition). Since the
BBN data we are using favours values of $\Delta N_{\rm eff}<1$ at
95\%~C.L., adding the BBN information slightly changes the best-fit
value for $\Delta N_{\rm eff}$ in the $\Lambda$CDM+$N_{\rm eff}$+BBN
analysis respect to the analysis of $\Lambda$CDM+$N_{\rm eff}$
(without the BBN information) and reduces the $1\sigma$ range from
$\Delta N_{\rm eff}\simeq 0.71^{+0.42}_{-0.38}$ (without BBN) to
$\Delta N_{\rm eff}\simeq 0.70^{+0.25}_{-0.30}$ (with BBN).

We find that the preferred lifetime is $\tau_{dec}\sim 10^3$ sec
which means decaying slightly before or during BBN. This can be 
understood because at any time $t$  the contribution of the decaying
matter and the dark radiation  produced in its decay to the energy density
of the universe,  parametrized as $\Delta N^{dec}_{\rm eff}$ and 
$\Delta N^{dr}_{\rm eff}$
respectively, can be approximated  by 
integrating Eqs.~(\ref{eq:rho_DM}) and (\ref{eq:rho_SR}):
\begin{eqnarray}
\Delta N^{dec}_{\rm eff}(t)
&=&
\left(\frac{8}{7}\right)\left(\frac{11}{4}\right)^{\frac{4}{3}}
\left(\frac{t}{t_0}\right)^{\frac{1}{2}}
\frac{\rho_{dec}(t_0)}{\rho_\gamma(t_0)} 
{\Large e}^{-\frac{t}{\tau_{dec}}}\,,
\label{eq:neffdec} \\
\Delta N^{dr}_{\rm eff}(t)
&=&
\left(\frac{8}{7}\right)\left(\frac{11}{4}\right)^{\frac{4}{3}}
\left(\frac{\tau_{dec}}{t_0}\right)^{\frac{1}{2}}
\frac{\rho_{dec}(t_0)}{\rho_\gamma(t_0)}
\left( \frac{\sqrt{\pi}}{2}{\rm Erf}\left[\sqrt\frac{t}{\tau_{dec}}\right]-
\sqrt\frac{t}{\tau_{dec}} {\Large e}^{-\frac{t}{\tau_{dec}}}
\right) \label{eq:neffdr}  \\
&&\xrightarrow[t \,\gg\, \tau_{dec}]{}
\left(\frac{8}{7}\right)\left(\frac{11}{4}\right)^{\frac{4}{3}}
\left(\frac{\tau_{dec}}{t_0}\right)^{\frac{1}{2}} \frac{\sqrt{\pi}}{2}
\frac{\rho_{dec}(t_0)}{\rho_\gamma(t_0)}\,,
\label{eq:neffdrlim} 
\end{eqnarray}
where we have assumed that the decay occurs in a radiation dominated
era, $a\propto \sqrt{t}$ and we have chosen for normalization 
$t_0=10^{-4}$s (up to the factor  $\frac{\sqrt{\pi}}{2}$
Eq.(\ref{eq:neffdrlim}) can be also obtained with the assumption of 
fast decay at $t=\tau_{dec}$~\cite{Fischler:2010xz}).

At CMB time  $\Delta N^{\rm CMB}_{\rm eff}$ is well approximated 
by Eq.~(\ref{eq:neffdrlim}) and the analysis constraints the product of
$\rho_{dec}(t_0)\times \sqrt{\tau_{dec}}$ so the 
density of decaying matter and its lifetime become strongly anticorrelated
(see also Fig.~\ref{fig:results2D_dec}). At BBN time 
\begin{eqnarray}
\Delta N_{\rm eff}^{\rm BBN}&=& 
\Delta N^{dec}_{\rm eff}(t_{\rm BBN})+
\Delta N^{dr}_{\rm eff}(t_{\rm BBN}) \nonumber \\
&=& \left(\frac{8}{7}\right)\left(\frac{11}{4}\right)^{4/3}
\left(\frac{\tau_{dec}}{t_0}\right)^{1/2}
\frac{\rho_{dec}(t_0)}{\rho_\gamma(t_0)}
\frac{\sqrt{\pi}}{2}{\rm Erf}\left[\sqrt\frac{t_{\rm BBN}}{\tau_{dec}}\right]
\;.
\label{eq:nbbn}
\end{eqnarray}
The present analysis favours  
$N_{\rm eff}^{\rm BBN}$ not very different from 
$\Delta N^{\rm CMB}_{\rm eff}$ and   
Eq.~(\ref{eq:nbbn}) and Eq.~(\ref{eq:neffdrlim}) give
comparable numerical results for $\tau_{dec}\sim t_{\rm BBN}$. 

From the one-dimensional probability plots, it can also be inferred
that the probability distributions for the parameter $\Delta N_{\rm
eff}$ in the $dec$M+BBN and in the model $\Lambda$CDM+$N_{\rm
eff}$+BBN are not equivalent. Besides the expected differences due to
the fact that in the  second scenario $N_{\rm eff}$ is constant 
in time while in
the first one it is not, there is a more subtle difference that is associated
with the prior distribution assumed for the parameters in the two
analysis.  In the $dec$M model, $N_{\rm eff}$ is a derived parameter
and it arises from a combination of two model parameters
$\tilde{\Omega}_{\rm dec}$ and $\Gamma_{\rm dec}$ 
as given in Eq.~(\ref{eq:neffdrlim}) and Eq.~(\ref{eq:nbbn}).  
Let us introduce an auxiliary variable $Z$, function of 
$\tilde{\Omega}_{\rm dec}$ 
and $\Gamma_{\rm dec}$, which takes values
along the $N_{\rm eff}$=cnt direction.  Using that the probability has
to be invariant under reparametrizations:
$\mathcal{P}(\tilde{\Omega}_{\rm dec},\Gamma_{\rm dec})
d\tilde{\Omega}_{\rm dec} d\Gamma_{\rm dec} =
\mathcal{\tilde{P}}(N_{\rm eff}, Z) d N_{\rm eff} d Z$.  The flat
prior for $\tilde{\Omega}_{\rm dec}$ and $\Gamma_{\rm dec}$ that we
are using in the analysis of the $dec$M scenario means that
$\mathcal{P}(\tilde{\Omega}_{\rm dec},\Gamma_{\rm dec})$=cnt.
Consequently, from Eqs.(\ref{eq:neffdec})--(\ref{eq:neffdrlim}), 
we find that the jacobian for the change of variables leads to 
$\mathcal{\tilde{P}}(N_{\rm eff},Z)\propto 1/N_{\rm eff}$.  
In other words, the effective prior in
$N_{\rm eff}$, that corresponds to having a flat prior in
$\{\tilde{\Omega}_{\rm dec},~\Gamma_{\rm dec}\}$, pushes $N_{\rm eff}$
to lower values.  Furthermore, since $H_{0}$, $\Omega_b h^2$, $\Omega_c
h^2$ and $n_s$ are correlated with the value of $\Delta N_{\rm eff}$,
this difference leads to the shift in the best-fit values for the
parameters $H_{0}$, $\Omega_b h^2$, $\Omega_c h^2$ and $n_s$ in the
two models seen in Tab.~\ref{tab:results} and in
Fig.~\ref{fig:results1D}. 

This also explains that the one-dimensional probability distribution
for the parameter $\Delta N_{\rm eff}^{\rm BBN}$ presents two peaks:
one around $\Delta N_{\rm eff}^{\rm BBN} \simeq 0$ and one around
$\Delta N_{\rm eff}^{\rm BBN} \simeq 0.5$. The first one is a
consequence of the $1/N_{\rm eff}$ ``initial'' probability
distribution in combination with the condition of vanishing dark
radiation before neutrino decoupling, Eq.~(\ref{eq:decneff}). The
second is induced by the condition of having an amount of dark 
radiation at BBN,
Eq.~(\ref{eq:bbnneff}).  Note, however, that the region below the
first peak is much smaller than the region below the second
peak. Indeed, the integral of $\int_{x_{\rm min}}^{+\infty} d x~p_{\rm
1-dim} (\Delta N^{\rm BBN}_{\rm eff}) \sim$~0.68, with $x_{\rm
min}\equiv 0.28$.  This means that even if the height of the first
peak is higher than the second one, a value of $\Delta N_{\rm
eff}^{\rm BBN} \gtrsim 0.28$ is favoured at roughly 68\%~C.L.. This is,
in turn, consistent with the best-fit value of the lifetime $\tau_{\rm
dec}$, that is around BBN time  
(see right panel in Fig.~\ref{fig:results2D_dec}).

In the right panel of Fig.~\ref{fig:results2D_dec} we present the
two-dimensional 68\% and 95\% C.L. credibility regions for the
parameters $\log(\tau_{dec}/\rm{s})$ and $\Delta N_{\rm eff}^{\rm
BBN}$ for the $dec$M+BBN analysis. In the figure, it is clearly visible 
that the regions are formed by  two connected ``islands'' around the two
favoured values of $\Delta N_{\rm eff}^{\rm BBN}$ . Also as expected the 
two parameters are anticorrelated: a
shorter lifetime is associated with a value of $\Delta N^{\rm
BBN}_{\rm eff}$ bigger than zero, while a lifetime bigger than $10^4$
will increase the value of $N^{\rm CMB}_{\rm eff}$ but not the value
of $N^{\rm BBN}_{\rm eff}$.

The left panel of Fig.~\ref{fig:results2D_dec} shows the
two-dimensional 68\% and 95\% C.L. credibility regions for the
parameters $\{\log(\rho_{dec}/\rho_\gamma)$ and
$\log(\tau_{dec}/\rm{s}) \}$. Here we also see the expected
anticorrelation between the total amount of decaying matter and its
lifetime required to produce a certain amount of dark radiation (see
Eq.(\ref{eq:neffdrlim})). The figure clearly shows how the best fit
values of the analysis correspond to a decay during the BBN time while
the contribution of the favoured range of $\rho_{dec}$ implies that its 
contribution to the dark matter at present times is totally negligible.

\section{Conclusions}
\label{sec:conclu} 
In this paper, we have explored the possibility that dark radiation is
not due to new relativistic particles but it is formed by SM neutrinos
whose density is increased by the decay of some heavy thermally
produced particle. With this aim we have performed a global analysis
of all the relevant cosmological data from the WMAP and SPT
collaborations, measurements of the Hubble constant at present time,
the results from high-redshift Type-I supernovae and the BAO scale.
We have also included the information on additional radiation at the
time of BBN as inferred from the relatively high $^4$He abundance
determination.  We have concluded that the inclusion of the BBN
information favours a decay that happens during BBN time with favoured
lifetime $\tau_{dec}$ between $10^2$ and $10^4$ seconds and leads to
similar values of $N_{\rm eff}$ at BBN and CMB times.  We have
discussed the difference in the analysis between this scenario and the
standard $\Lambda$CDM+$N_{\rm eff}$ associated with the different
physical assumptions on the initial probability
distribution of the model parameters. 

The discussion of a specific theoretical model in which this scenario is
implemented is beyond the scope of this paper.  Different possibilities
have been suggested in the literature in the context of a $dec$M that
decays into radiation and that can be applied with some extents also
to our case of decay into neutrinos.  We refer to
Refs.~\cite{Fischler:2010xz,Ichikawa:2007jv} for different examples.


\section*{Acknowledgments}
This work is  supported by USA-NSF grant PHY-09-6739, by
CUR Generalitat de Catalunya grant 2009SGR502 by MICINN FPA2010-20807 
and consolider-ingenio 2010 program grants CUP (CSD-2008-0037) and CPAN,
and by EU grant 
FP7 ITN INVISIBLES (Marie Curie Actions PITN-GA-2011-289442).
J.S. acknowledges support from the Wisconsin IceCube Particle
Astrophysics Center (WIPAC) and U. S. Department of Energy under the
contract DE-FG-02-95ER40896.

\clearpage 

\bibliographystyle{JHEP}
\bibliography{decaying_final}


\end{document}